\documentclass[letterpaper,twocolumn,prx,aps,showpacs,superscriptaddress,floatfix]{revtex4-1}
\usepackage[latin1]{inputenc}
\usepackage{bm}
\usepackage[usenames]{color}
\usepackage{multirow}
\usepackage{amssymb}
\usepackage{amsbsy}
\usepackage{amsmath}
\usepackage{stmaryrd}
\usepackage{graphicx}
\usepackage{epsfig}
\usepackage{placeins}
\usepackage{ulem}
\usepackage{bbm}
\usepackage{braket}
\usepackage[colorlinks,linkcolor=blue,citecolor=blue,urlcolor=blue]{hyperref}
\usepackage{blindtext}

\makeatletter

\begin{document}
\title{Magnetization dynamics in clean and disordered spin-1 XXZ chains}

\author{Jonas Richter}
\email{jonasrichter@uos.de}
\affiliation{Department of Physics, University of Osnabr\"uck, D-49069 Osnabr\"uck, Germany}

\author{Niklas Casper}
\affiliation{Institute for Theoretical Physics, Technical University 
Braunschweig, D-38106 Braunschweig, Germany}

\author{Wolfram Brenig}
\affiliation{Institute for Theoretical Physics, Technical University 
Braunschweig, D-38106 Braunschweig, Germany}

\author{Robin Steinigeweg}
\email{rsteinig@uos.de}
\affiliation{Department of Physics, University of Osnabr\"uck, D-49069 Osnabr\"uck, Germany}

\date{\today}

\begin{abstract}

We study spin transport in the one-dimensional anisotropic $S = 1$ Heisenberg 
model. Particular emphasis is given to dynamics at infinite temperature, where  
current autocorrelations and spatio-temporal correlation functions are obtained 
by means of an efficient pure-state approach based on the concept of 
typicality. Our comprehensive numerical analysis unveils that high-temperature 
spin transport is diffusive in the easy-axis regime for strong exchange 
anisotropies. This finding is based on the combination of numerous 
signatures, such as (i) Gaussian spreading of correlations, (ii) 
a time-independent diffusion coefficient, (iii) power-law decay of 
equal-site correlations, (iv) exponentially decaying long-wavelength modes, and 
(v) Lorentzian line shapes of the dynamical structure factor. Moreover, we 
provide evidence that some of these signatures are not exclusively restricted to 
the infinite-temperature limit, but can persist at lower temperatures as well, 
where we complement our results by additional quantum Monte Carlo simulations 
of large systems. In contrast to the easy-axis regime, we show that in the case 
of an isotropic chain, the signatures (i) - (v) are much less pronounced or even 
entirely absent, suggesting the existence of anomalous spin transport despite 
the nonintegrability of the model. Eventually, upon introducing a random 
on-site magnetic field, we observe a breakdown of diffusion and distinctly 
slower dynamics. In particular, our results exhibit qualitative similarities to 
disordered spin-$1/2$ chains and might be consistent with the onset of 
many-body localization in the $S = 1$ model for sufficiently strong disorder.  

\end{abstract}

\maketitle

\section{Introduction}

Fundamental aspects about the emergence of thermodynamic behavior in closed 
quantum many-body systems have recently attracted rejuvenated interest 
\cite{Polkovnikov2011, Gogolin2016, Dallesio2016}. While there has been 
immense progress due to, e.g., theoretical concepts such as the eigenstate 
thermalization hypothesis \cite{deutsch1991, srednicki1994, rigol2005}, 
large-scale numerical simulations \cite{schollwoeck20052011}, as well as the 
advance of new experimental platforms \cite{Langen2015, Blatt2012}, there are 
still challenging problems calling for a deeper understanding. For instance, a 
particularly intriguing question is whether or not conventional hydrodynamic 
transport, i.e., diffusion, can arise in isolated quantum systems undergoing 
solely unitary time evolution \cite{buchanan2005}. 

Intimately related to this question is the distinction between integrable and 
nonintegrable models. On the one hand, integrable models are characterized by a 
macroscopic number of (quasi)local conservation laws which can lead to anomalous 
thermalization \cite{essler2016, vidmar2016} and ballistic transport 
\cite{zotos1997, prosen2013, heidrichmeisner2003}. As a consequence, diffusion 
is generally not expected to occur in these systems. Nevertheless, while 
subleading diffusive corrections have been 
established within the framework of generalized hydrodynamics 
\cite{Alvaredo2016, Bertini2016, DeNardis2018}, there is 
also clear numerical evidence for diffusive transport in certain
integrable models and parameter regimes \cite{Sirker2011, znidaric2011, 
steinigeweg2011_1, karrasch2014_2, Steinigeweg2017, Ljubotina2017, prosen2012, 
karrasch2017, steinigeweg2017_2}. 

On the other hand, integrability is rather the exception than the rule and can 
be broken in numerous ways, e.g., spin-phonon coupling \cite{Chernyshev2016, 
jansen2019}, long-range interactions \cite{Hazzard2014, Kloss2019}, impurities 
\cite{Metavitsiadis2010, Brenes2018}, or disorder \cite{Herbrych2013}. For such 
nonintegrable models, Drude 
weights are expected to vanish \cite{heidrichmeisner2003} and diffusion might 
emerge, e.g., due to quantum chaos \cite{Borgonovi2016}. Although much progress 
in understanding the 
emergence of diffusive hydrodynamics has been recently made in the study of 
random unitary 
circuit models \cite{Keyserlingk2018, Nahum2018, Khemani2018}, observations of 
genuine diffusion in realistic systems are comparatively rare \cite{Michel2005, 
Monasterio2005, karrasch2014_2, Medenjak2017, Richter2018_1, Richter2019_2}. In 
particular, it is still an open question if diffusion is indeed a generic 
feature for all integrability-breaking perturbations (note that counterexamples 
have been proposed \cite{Brenes2018}). Moreover, answering this question is also very 
challenging due to the tremendous numerical requirements which arise in the 
study of transport in quantum many-body systems, such as the exponential growth 
of the Hilbert-space dimension as well as the necessity to study long time 
scales.   

In this context, we consider yet another nontrivial way to break integrability, 
i.e, the consideration of a larger spin quantum number $S > 1/2$ 
\cite{Haldane1983, Mikeska2004}. In 
particular, we study spin transport in the one-dimensional $S = 1$ XXZ model, 
using an efficient numerical approach based on the concept of quantum 
typicality  \cite{lloydPhd, Gemmer2004, Popescu2006, Goldstein2006, 
Reimann2007, Hams2000, 
iitaka2003, sugiura2013, elsayed2013, monnai2014, steinigeweg2014}. Summarizing 
our main results, we unveil that high-temperature spin transport is diffusive in 
the easy-axis regime of large anisotropies. This finding is based on the 
combination of numerous signatures, such as (i) Gaussian spreading of  
correlations, (ii) a time-independent diffusion coefficient, (iii) power-law 
decay of equal-site correlations, (iv) exponentially decaying 
long-wavelength modes, and (v) Lorentzian line shapes of the dynamical 
structure 
factor. Moreover, we provide evidence that some of these signatures are not 
exclusively 
restricted to the infinite-temperature limit, but can persist at lower 
temperatures as well, where we complement our results by additional quantum 
Monte Carlo simulations of large systems. In contrast to the easy-axis regime, 
we show that in the case of an isotropic chain, the signatures (i) - (v) are 
much less pronounced or even entirely absent, suggesting the existence of 
anomalous spin transport despite the nonintegrability of the model. Eventually, 
upon introducing a random on-site magnetic field, we observe a breakdown of 
diffusion and distinctly slower dynamics. In particular, our results exhibit 
qualitative similarities to disordered spin-$1/2$ chains and might be 
consistent with the onset of many-body localization in the $S = 1$ model for 
sufficiently strong disorder.  

This paper is structured as follows. We introduce the model in 
Sec.\ \ref{Sec::Model} and define the observables in 
Sec.\ \ref{Sec::Observables}. In Sec.\ \ref{Sec::Numerics} we explain our 
numerical approach, and we present our results in Sec.\ref{Sec::Results}. We 
conclude and summarize in Sec.\ \ref{Sec::Conclusion}.  


\section{Model}\label{Sec::Model}

We study the one-dimensional $S = 1$ XXZ model with periodic boundary 
conditions, described by the Hamiltonian
\begin{equation}\label{Eq::Ham1}
 \mathcal{H} =  J\sum_{l=1}^L \left(S_l^x S_{l+1}^x + S_l^y S_{l+1}^y 
 + \Delta S_l^z S_{l+1}^z\right)\ ,
\end{equation}
where the $S_l^{x,y,z}$ are spin-$1$ operators at lattice site $l$, $J=1$ 
denotes the antiferromagnetic exchange constant (and sets the energy scale 
throughout this paper), $L$ is the number of sites, and $\Delta > 0$ is the 
anisotropy in $z$ direction. In contrast to its spin-$1/2$ counterpart, the 
Hamiltonian \eqref{Eq::Ham1} is nonintegrable for $S =1$ \cite{Mikeska2004, 
Piroli2016}. 

The spin-$1$ chain \eqref{Eq::Ham1} is a fundamental model of low-dimensional 
quantum magnetism and is realized to good quality in numerous materials. As a 
consequence, its thermodynamic and its dynamical properties have been 
intensively scrutinized by theoretical \cite{Mikeska2004, Deisz1990, White1993, 
Karadamoglou2004, GrossjohannDiss, Becker2017, Capponi2019} and experimental 
techniques 
\cite{Regnault1994, Orendac1995, Honda1998, Kenzelmann2002, Takigawa1996, 
Sologubenko2003}. Moreover, various modifications to the bare Hamiltonian 
\eqref{Eq::Ham1} have been explored as well, such as, e.g., single-ion 
anisotropy and external magnetic fields \cite{Affleck1991, Rahnavard2015, 
Furuya2011, Herbrych2016, Lange2018}. While experiments have reported on the 
existence of diffusive spin and energy transport in spin-$1$ compounds 
\cite{Takigawa1996, Sologubenko2003}, theoretical studies have given 
contradictory results and argued for diffusive \cite{Karadamoglou2004, 
Sachdev1997, Steinigeweg2010} but also ballistic transport \cite{Fujimoto1999, 
Konik2003}. 
In this context, it is important to note that Refs.\ 
 \cite{Sachdev1997, Fujimoto1999, Konik2003} have in fact considered the 
non-linear sigma model as the effective low-energy description of Eq.\ 
\eqref{Eq::Ham1}, where additional conservation laws might have an impact on 
the transport properties.
 
While the focus of this paper is on spin $S = 1$, it is instructive to 
briefly recap the nature of spin dynamics in the integrable $S = 1/2$ version 
of Eq.\ \eqref{Eq::Ham1}. On the one hand, for $\Delta < 1$, the spin-$1/2$ 
chain features a finite Drude weight, i.e., ballistic transport \cite{zotos1999, 
prosen2013, urichuk2019}. On the 
other hand, for $\Delta > 1$, the Drude weight vanishes and clean 
signatures of diffusion have been observed \cite{Sirker2011, znidaric2011, 
steinigeweg2011_1, karrasch2014_2, Steinigeweg2017, Ljubotina2017}. While the 
situation is 
arguably most controversial for $\Delta = 1$, recent works advocate the 
presence of superdiffusion at the isotropic point at high temperatures 
\cite{Ljubotina2017, Gopalakrishnan2018, Richter2019}. In 
this context, it is an intriguing question if normal diffusion generically 
occurs in the one-dimensional XXZ 
model (i.e. for all $\Delta$) upon considering the larger spin quantum number 
$S=1$, both at infinite and also finite temperatures (see also Ref.\ 
\cite{DeNardis2019}). We explore this question in Secs.\ \ref{Sec::Results::CH} 
and \ref{Sec::Results::CL}. 

In Sec.\ \ref{Sec::Results::DL}, we additionally study the spin-$1$ XXZ chain 
in the presence of a random magnetic field, i.e., the 
Hamiltonian \eqref{Eq::Ham1} is modified according to  
\begin{equation}\label{Eq::Ham2}
 \mathcal{H} =  J\sum_{l=1}^L \left(S_l^x S_{l+1}^x + S_l^y S_{l+1}^y 
 + \Delta S_l^z S_{l+1}^z + h_l S_l^z \right)\ ,
\end{equation}
where the on-site magnetic fields $h_l \in [-W,W]$ are drawn at random from a 
uniform distribution, with $W\geq 0$ setting the magnitude of disorder. 

Once again, let us briefly reiterate the case of $S = 1/2$. In fact, the 
disordered spin-$1/2$ Heisenberg chain is a central model to study the 
disorder-driven transition between a thermal phase ($W < 
W_c$) and a many-body-localized (MBL) phase ($W > W_c$), where $W_c$ is a 
critical disorder strength \cite{basko2006, nandkishore2015}. This MBL phase is 
characterized by, e.g., a vanishing dc conductivity \cite{Berkelbach2010, 
Gopalakrishnan2015}, area-law entanglement of eigenstates \cite{luitz2015, 
Bauer2013}, the emergence of a set of local integrals of motion \cite{Huse2014, 
Serbyn2013}, as well as the logarithmic growth of entanglement with time 
\cite{Znidaric2008, Bardarson2012}.
While ground-state properties of disordered spin-$1$ systems have been studied 
before \cite{Monthus1998, Refael2007}, their dynamics has remained largely 
unexplored. Therefore, the present paper attempts to elucidate the effect of 
disorder on spin dynamics in the anisotropic spin-$1$ chain.


\section{Observables}\label{Sec::Observables}

Let us now introduce the quantities which are studied in this paper. In 
particular, we discuss how diffusive transport can be detected based on these 
quantities. 

\subsection{Current autocorrelations and 
transport coefficients}\label{Sec::Observables::Current}

Since total magnetization is conserved for all choices of $\Delta$ and $W$, the 
spin current $j = \sum_l j_l$ is well defined via a lattice continuity 
equation, $\partial_t S_l^z = i[\mathcal{H},S_l^z] = j_{l-1} - j_{l}$, and 
takes on the form \cite{heidrichmeisner2007}
\begin{equation}\label{Eq::Current}
  j = J \sum_{l=1}^L \left(S_l^x S_{l+1}^y - S_l^y S_{l+1}^x \right)\ . 
\end{equation}
Within linear response theory, transport properties are related to 
current-current correlation functions evaluated in equilibrium, 
\begin{equation}\label{Eq::CurCur}
 \langle j(t) j \rangle = \frac{\text{Tr}[e^{-\beta {\cal H}}j(t)j]}{{\cal Z}}\ ,  
\end{equation}
where the time argument has to be understood in the Heisenberg picture 
$j(t) = e^{i\mathcal{H}t} j e^{-i\mathcal{H}t}$, $\beta = 1/T$ denotes the 
inverse temperature, and ${\cal Z}=\text{Tr}[e^{-\beta {\cal H}}]$ is the 
partition function. For instance, integration of $\langle j(t) j \rangle$ 
yields the time-dependent diffusion coefficient $D(t)$ \cite{Steinigeweg2009},
\begin{equation}\label{Eq::Dt}
D(t) = \frac{1}{\chi} \int_0^t \frac{\text{Re}\ \langle j(t') j \rangle}{L}\  \text{d}t'\ ,
\end{equation}
where $\chi = \lim_{q\to 0} \langle S_q^z S_{-q}^z \rangle$ denotes the 
isothermal spin susceptibility. In the case of diffusion, one expects that the 
current autocorrelation eventually decays to zero, such that $D(t)$ saturates 
at a constant plateau $D(t > \tau)\approx D$ for times $t$ above the mean-free 
time $\tau$. Generally, however, it is important to note that 
$D(t)$ does not distinguish between transport channels with different 
behavior \cite{Sirker2011}. 

Furthermore, the frequency-dependent spin conductivity $\sigma(\omega)$ follows 
from a Fourier transform of the current-current correlation function, 
\begin{equation}
  \text{Re}\ \sigma(\omega) = \frac{1-e^{-\beta\omega}}{\omega L}\ \text{Re} \int\limits_0^\infty  
e^{i\omega t}\ \langle j(t) j \rangle\ \text{d}t\ , \label{Eq::SigOm}
\end{equation} 
and is usually decomposed into a $\delta$ function at $\omega = 
0$ and a regular part at $\omega\neq 0$, 
\begin{equation}
 \text{Re}\ \sigma(\omega) = {\cal D} \delta(\omega) + \sigma_\text{reg}(\omega)\ ,
\end{equation}
where ${\cal D}$ is the so-called Drude weight. If transport is 
diffusive, 
we have ${\cal D} = 0$ and there is a well-behaved dc conductivity 
$\sigma_\text{dc} = \lim_{\omega \to 0} \sigma_\text{reg}(\omega)$. Moreover, 
using an Einstein relation, this dc conductivity is connected to the diffusion 
constant according to \cite{Steinigeweg2009}, 
\begin{equation}
 D = \frac{\sigma_\text{dc}}{\chi}\ . 
\end{equation}

\subsection{Spin 
density correlations}\label{Sec::Observables::Densities}

In addition to current dynamics, we also study the dynamics of spatio-temporal 
correlation functions $C_{l,l'}(t)$ defined as, 
\begin{equation}\label{Eq::Cll}
C_{l,l'}(t) = \langle S_l^z(t) S_{l'}^z \rangle = \frac{\text{Tr}[e^{-\beta 
{\cal H}}S_l^z(t)S_{l'}^z]}{{\cal Z}}\\ . 
\end{equation}
For the particular case of $\beta \to 0$, these correlations realize 
a $\delta$-peak profile at time 
$t = 0$, or in other words, spins at different lattice sites are uncorrelated 
at infinite temperature,  
\begin{equation}\label{Eq::InitProfile}
 C_{l,l'}(t= 0) = \begin{cases}
                \chi > 0, &l = l' \\
                0, &l \neq l'
               \end{cases}\ , 
\end{equation}
with $\chi = 2/3$ for $\beta = 0$ and $S = 1$. For times $t > 0$, however, 
correlations 
start to build up and the initial $\delta$ peak will spread over the system. 
Specifically, in the case of diffusion, this spreading 
yields a Gaussian density profile \cite{Steinigeweg2017, Richter2018_1}, 
\begin{equation}
 C_{l,l'}(t) \propto \exp\left[-\frac{(l-l')^2}{2\Sigma(t)^2}\right]\ ,   
\end{equation}
where the spatial variance $\Sigma(t)^2$ is generally given by 
\begin{equation}\label{Eq::SigDense}
 \Sigma(t)^2 = \sum_{l=1}^L l^2\ \delta C_{l,l'}(t) - \left(\sum_{l=1}^L l\ 
\delta 
C_{l,l'}(t) \right)^2\ ,
\end{equation}
with $\delta C_{l,l'}(t) = C_{l,l'}(t)/\chi$ and $\sum_l \delta C_{l,l'}(t) = 
1$. Due to 
continuity, this spatial variance is also related to the already mentioned 
diffusion coefficient \cite{steinigeweg2009_2, yan2015, luitz2017},  
\begin{equation}\label{Eq::Sigt}
 \frac{\text{d}}{\text{d}t}\Sigma(t)^2 = 2D(t)\ . 
\end{equation}
Note that Eq.\ \eqref{Eq::Sigt} does not require specific 
assumptions on the microscopic Hamiltonian ${\cal H}$, apart from $[{\cal 
H}, \sum_l S_l^z ] = 0$ and the system being translational invariant.
Given a diffusive process, i.e., $D(t) = D = \text{const.}$, 
it then follows that $\Sigma(t)^2 \propto t$. Moreover, this particular scaling 
of $\Sigma(t)^2$ also implies that the equal-site correlation $C_{l,l'=l}(t)$ 
decays as a power-law 
\begin{equation}\label{Eq::PL}
 C_{l,l'=l}(t) \propto t^{-1/2}\ . 
\end{equation}

Starting from the real-space correlations in Eq.\ \eqref{Eq::Cll}, the 
respective correlation functions in momentum space follow from a lattice 
Fourier transform according to \cite{Fabricius1997}
 \begin{align}
  C_q(t) = \langle S_q^z(t) S_{-q}^z \rangle &= \frac{1}{L}\sum_{l,l'=1}^L 
	    e^{iql} e^{-iql'} \langle S_l^z(t) S_{l'}^z 
\rangle \label{Eq::LatFou} \\ 
 	 &= \sum_{l=1}^L e^{iql} \langle S_{l'+l}^z(t) S_{l'}^z \rangle\ ,  
\label{Eq::Cq2}
 \end{align}
where the discrete momenta $q$ are defined as usual, 
$q = 2\pi k /L$, $k = 0, 1, \dots, L-1$. Note that Eq.\ \eqref{Eq::Cq2} is 
strictly valid only for translational invariant systems ($W = 0$), but might 
also hold approximately for $W > 0$ if the $C_{l,l'}(t)$ are averaged over 
sufficiently many disorder realizations, cf.\ Sec.\ \ref{Sec::Av}. In momentum 
space, diffusion can be 
characterized by the existence of a hydrodynamic regime where long-wavelength 
modes exhibit an exponential decay, 
\begin{equation}\label{Eq::Expo}
 C_q(t) \propto e^{-\tilde{q}^2Dt}\ ,
\end{equation}
with $\tilde{q}^2 = 2[1 - \cos(q)] \approx q^2$ for small $q$.

Moreover, another Fourier transform from the time to the frequency domain 
yields the so-called dynamical structure factor $C_q(\omega)$, 
 \begin{equation}\label{Eq::Sqw}
  C_q(\omega) = \int_{-\infty}^\infty e^{i\omega t} C_q(t)\ \text{d}t\ . 
 \end{equation}
As a direct consequence of the exponentials in Eq.~\eqref{Eq::Expo}, diffusive 
transport reflects itself in a Lorentzian line shape of $C_q(\omega)$, 
 \begin{equation}\label{Eq::Lorentz}
  C_q(\omega) \propto \frac{1}{\omega^2 + \tilde{q}^4 D^2}\ , 
 \end{equation}
for sufficiently long wavelengths. 
  

\section{Numerical approach}\label{Sec::Numerics}

\subsection{Dynamical quantum typicality}\label{Sec::DQT}

Loosely speaking, the concept of dynamical quantum typicality (DQT) states that 
a single pure quantum state can have the same properties as the statistical 
ensemble \cite{lloydPhd, Gemmer2004, Popescu2006, Goldstein2006, 
Reimann2007, Hams2000, iitaka2003, sugiura2013, elsayed2013, monnai2014, 
steinigeweg2014}. In practice, this fact can be exploited in order to replace 
the trace in Eq.\ \eqref{Eq::CurCur} by a simple scalar product with two 
auxiliary pure states $\ket{\varphi_\beta(t)}$, $\ket{\psi_\beta(t)}$ such that 
the current autocorrelation takes on the form \cite{elsayed2013, 
steinigeweg2014, Steinigeweg2015}
\begin{equation}\label{Eq::Typ}
 \text{Re}\ \langle j(t) j \rangle = \frac{\text{Re}\ \bra{\varphi_\beta(t)} j 
\ket{\psi_\beta(t)}}
 {\braket{\varphi_\beta(0)|\varphi_\beta(0)}} + \epsilon\ , 
\end{equation}
with $\ket{\varphi_\beta(0)} = e^{-\beta {\cal H}/2}\ket{\varphi}$, 
$\ket{\psi_\beta(0)} = j\ e^{-\beta {\cal H}/2}\ket{\varphi}$, and
\begin{equation}\label{Eq::Varphi}
 \ket{\varphi} = \sum_{k=1}^d c_k \ket{\phi_k}\ . 
\end{equation}
Here, the reference pure state $\ket{\varphi}$ is prepared according to the 
unitary invariant Haar measure \cite{bartsch2009}, i.e., the complex 
coefficients $c_k$ 
are randomly drawn from a Gaussian distribution with zero mean. The states 
$\ket{\phi_k}$ denote a complete set of basis vectors of the Hilbert space, 
e.g., the Ising basis. If not stated otherwise, we always consider the full 
Hilbert space with dimension $d = 3^L$. 

Importantly, the statistical error $\epsilon = \epsilon(\ket{\varphi})$ in Eq.\ 
\eqref{Eq::Typ} scales as $\epsilon \propto 1/\sqrt{d_\text{eff}}$, where 
$d_\text{eff} = {\cal Z}/e^{-\beta E_0}$
is an effective dimension and $E_0$ is the ground-state energy of ${\cal H}$ 
\cite{Hams2000, bartsch2009, elsayed2013, steinigeweg2014, Steinigeweg2015}. 
Thus, $\epsilon$ decreases exponentially with increasing system size 
and, particularly for $\beta \to 0$, the typicality approximation becomes very 
accurate already for moderate values of $L$ \cite{steinigeweg2014, 
Richter2019}. 

Completely analogous to current autocorrelations, the spatio-temporal 
correlations $C_{l,l'}(t)$ for $\beta \geq 0$ can be obtained by means of a 
pure-state approach as well, simply by replacing $j$ with $S_l^z$ (or 
$S_{l'}^z$) in and below Eq.\ \eqref{Eq::Typ}. However, in the limit $\beta \to 
0$, it is even possible to calculate $C_{l,l'}(t)$ from just \textit{one} 
auxiliary state \cite{Richter2019_2, Richter2019_1} (see also 
Appendix~\ref{Sec::Appendix::Typ}), 
\begin{equation}\label{Eq::Typ2}
 \text{Re}\ C_{l,l'}(t) = \text{Re}\ \bra{\tilde{\psi}(t)} S_l^z 
\ket{\tilde{\psi}(t)} + \epsilon\ , 
\end{equation}
where $\ket{\tilde{\psi}(t=0)}$ is constructed according to
\begin{equation}\label{Eq::Psi}
 \ket{\tilde{\psi}(0)} = \frac{\sqrt{S_{l'}^z+1} 
\ket{\varphi}}{\sqrt{\braket{\varphi|\varphi}}}\ , 
\end{equation}
with $\ket{\varphi}$ again being randomly drawn, cf.\ Eq.\ \eqref{Eq::Varphi}. 
It is 
important to note that the operator $S_{l'}^z+1$ (i) only has nonnegative 
eigenvalues and (ii) is diagonal in the Ising basis. Therefore, the application 
of the square root operation is well-defined and rather straightforward 
\cite{Richter2019_2}. 

\subsection{Pure-state propagation}\label{Sec::PSP}

In Eqs.\ \eqref{Eq::Typ} and \eqref{Eq::Typ2}, the time argument is interpreted 
as a property of the pure states and not of the operators anymore, 
 \begin{equation}
  \ket{\psi(t)} = e^{-i{\cal H}t} \ket{\psi(0)}\ . 
 \end{equation}
Compared to standard exact diagonalization (ED), the main advantage of the 
pure-state approach stems from the fact that this time evolution can be 
conveniently evaluated by means of an iterative forward propagation, 
$\ket{\psi(t+\delta t)} = e^{-i\mathcal{H}\delta t} \ket{\psi(t)}$. Similarly, 
the action of $e^{-\beta {\cal H}/2}$ can be generated by a forward propagation 
as well, but now in imaginary time. While there exist various sophisticated 
methods such as Trotter decompositions \cite{deReadt2006}, Chebyshev expansions 
\cite{dobrovitski2003, weisse2006}, as well as Krylov-subspace techniques 
\cite{varma2017}, we here apply a fourth-order 
Runge-Kutta algorithm where the discrete time step $\delta t$ is always chosen 
short enough to guarantee negligible numerical errors \cite{elsayed2013, 
steinigeweg2014}. Since the involved operators usually exhibit a sparse matrix 
representation, the matrix-vector multiplications within this Runge-Kutta 
scheme can be implemented memory efficient, and we can treat Hilbert-space 
dimensions significantly larger compared to ED. (As an example, $d = 
3^L \approx 4\cdot10^8$ for $L = 18$. Note that this Hilbert-space 
dimension would correspond to $L \approx 29$ in the case of $S = 1/2$.) 

Eventually, concerning the Fourier transforms in Eqs.\ \eqref{Eq::SigOm} and 
\eqref{Eq::Sqw}, let us note that the integrals can in practice be evaluated 
only up to a finite cutoff time $t_\text{max} < \infty$, giving rise to a finite 
frequency resolution $\delta \omega = \pi/t_\text{max}$. 
\begin{figure}[tb]
  \centering 
  \includegraphics[width=0.9\columnwidth]{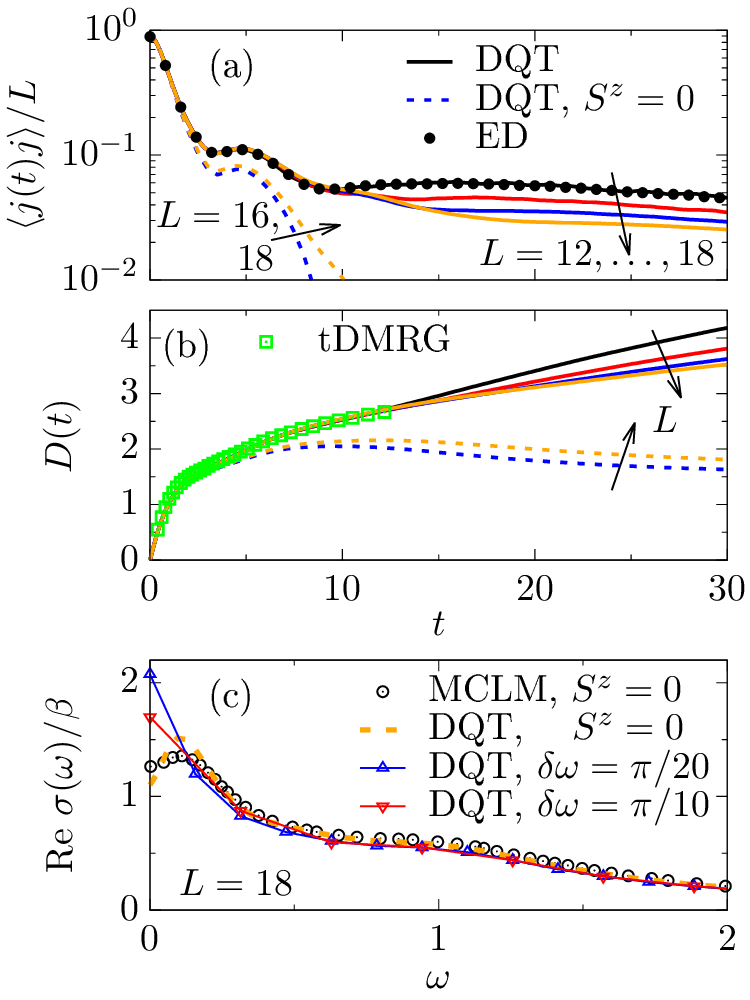}
  \caption{(Color online) (a) Current autocorrelation $\langle j(t) j 
\rangle/L$ obtained for different system sizes $L = 12, 14, 16, 18$ by ED and 
DQT. As a comparison, we also show data for $L = 16, 18$ calculated 
in the $S^z = 0$ subsector only. (b) Corresponding diffusion coefficient $D(t)$, 
cf.\ Eq.\ \eqref{Eq::Dt}. Rescaled data from time-dependent 
density matrix renormalization group (tDMRG) calculations for $T = 10$ are 
depicted \cite{DeNardis2019, NotetDMRG}. (c) Conductivity $\sigma(\omega)$ 
calculated according to Eq.\ \eqref{Eq::SigOm} for frequency resolutions 
$\delta \omega = \pi/10$, $\pi/20$, and $\pi/100$ ($S^z = 0$). As a comparison, 
we depict data obtained by the microcanonical Lanczos method (MCLM) from Ref.\ 
\cite{Karadamoglou2004}. The other parameters are $\Delta = 1$ and $\beta = 0$.}
  \label{Fig1}
\end{figure}
\begin{figure}[tb]
  \centering 
  \includegraphics[width=0.9\columnwidth]{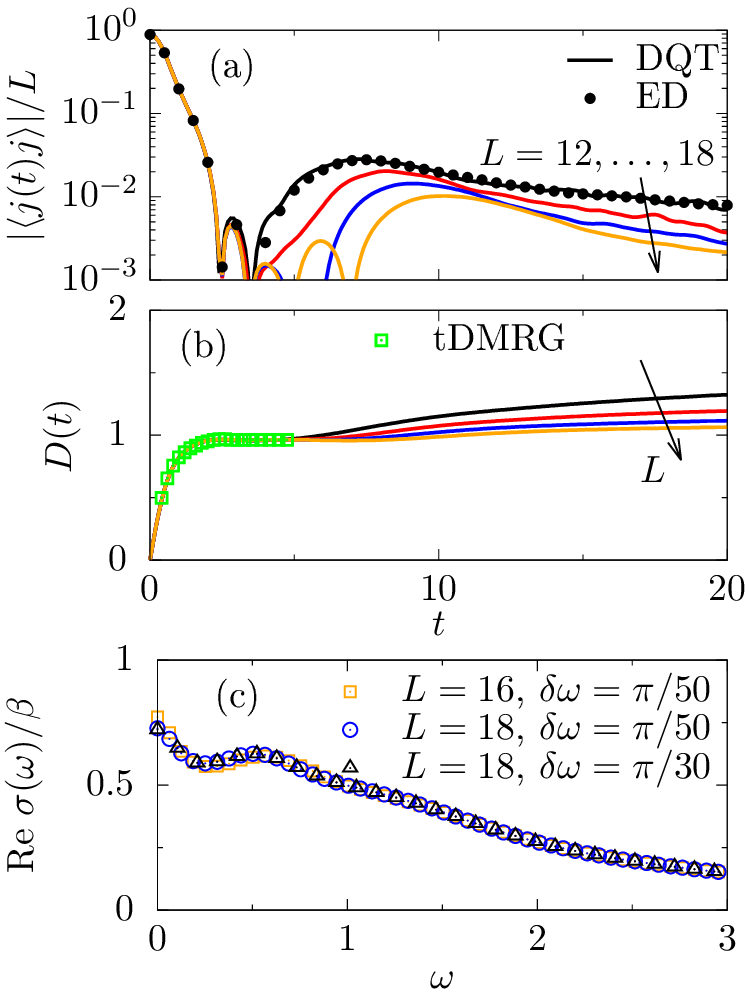}
  \caption{(Color online) Analogous data as in Fig.\ \ref{Fig1}, but now for 
    a larger anisotropy $\Delta = 1.5$. Note that we plot the absolute value 
  $|\langle j(t) j \rangle|/L$ in (a) for better visibility. All  
calculations are performed in the full Hilbert space.}
  \label{Fig2}
\end{figure}

\subsection{Averaging}\label{Sec::Av}

In this paper, we have to differentiate between two possible types of averaging. 
On the one hand, our numerical approach is based on the construction of the 
pure state $\ket{\varphi}$, cf.~\eqref{Eq::Varphi}, comprising the random 
coefficients $c_k$. Although the statistical error $\epsilon(\ket{\varphi})$ of 
the typicality approximation is rather small for large $L$, the remaining error 
can be reduced even further by averaging over $N_S$ different instances of the 
$c_k$. While such a procedure is usually unnecessary for $\beta \to 0$ 
($N_S = 1$), it can be beneficial for temperature regimes $T\lesssim J$ 
\cite{iitaka2003, Rousochatzakis2018}. 

On the other hand, in the case of a disordered model with $W > 0$, all results 
naturally depend on the specific configuration of the random magnetic fields 
$h_l$. In order to obtain reliable results, we therefore routinely perform an 
averaging over a sufficiently large number $N$, 
\begin{equation}
 C_{l,l'}(t) = \frac{1}{N} \sum_{n=1}^N C_{l,l'}^{(n)}(t)\ , 
\end{equation}
where each $C_{l,l'}^{(n)}(t)$ is evaluated for a different random 
configuration of the $h_l$. 

A useful measure, both for sampling over initial states as well as over 
disorder configurations, is the variance of sample-to-sample fluctuations,  
\begin{equation}
 \Delta C_{l,l'}(t) = \sum_{n=1}^{N_{(S)}} 
\frac{[C_{l,l'}^{(n)}(t)]^2}{N_{(S)}} - \left(\sum_{n=1}^{N_{(S)}} 
\frac{C_{l,l'}^{(n)}(t)}{N_{(S)}} \right)^2\ , 
\end{equation}
which will typically increase for lower temperatures $T$ and stronger disorder 
$W$. The value of $N_{(S)}$ has to be chosen in such a way that the error of 
the mean $\sqrt{\Delta C_{l,l'}(t)/N_{(S)}}$ remains small in all 
cases. The above reasoning of course not only applies to the 
spatio-temporal correlations $C_{l,l'}(t)$, but also to the current 
autocorrelation $\langle j(t) j \rangle$. 

\subsection{Quantum Monte Carlo}

The quantum Monte Carlo (QMC) method is based on the stochastic series 
expansion (SSE) \cite{Sandvik1992, Sandvik1999, Syljuasen2002}, which uses 
importance 
sampling of the high-temperature series expansion of the partition function 
with a truncation of the sum to order $M$,
\begin{equation}\label{eq:partition}
{\cal Z}=\sum_{\alpha}\sum_{S_{M}}\frac{(-\beta)^{n}(M-n)!}{M!}\Braket{
\alpha|\prod_ {p=1}^{M}{\cal H}_{a_{p},b_{p}}|\alpha} \, ,
\end{equation}
where $a_p = 1,2$ indicates diagonal ${\cal 
H}_{1,b}=J \Delta S_{i(b)}^{z}S_{j(b)}^{z} + C$ or off-diagonal ${\cal 
H}_{2,b}=J(S_{i(b)}^{+}S_{j(b)}^{-}+ \text{h.c.})/2$ operators on bond $b$. The 
constant $C$ is chosen such that all diagonal weights are positive 
\cite{Syljuasen2002}, 
$\Ket{\alpha}=\Ket{S_{1}^{z},\ldots,S_{L}^{z}}$ refers to the $S^{z}$ 
basis, and $S_{M}=[a_{1},b_{1}][a_{2},b_{2}]\ldots[a_{M},b_{M}]$ is an index 
for the operator string $\prod_{p=1}^{M}{\cal H}_{a_{p},b_{p}}$. This string is 
Metropolis sampled, using two types of updates, (i) diagonal updates which 
change the number of diagonal operators ${\cal H}_{1,b_{p}}$ in the operator 
string and (ii) loop updates which change the type of operators 
${\cal H}_{1,b_{p}}\leftrightarrow {\cal H}_{2,b_{p}}$. For bipartite lattices 
the loop update comprises an even number of off-diagonal operators 
${\cal H}_{2,b_{p}}$, ensuring positivity of the transition probabilities.
\begin{figure}[tb]
  \centering 
  \includegraphics[width=0.9\columnwidth]{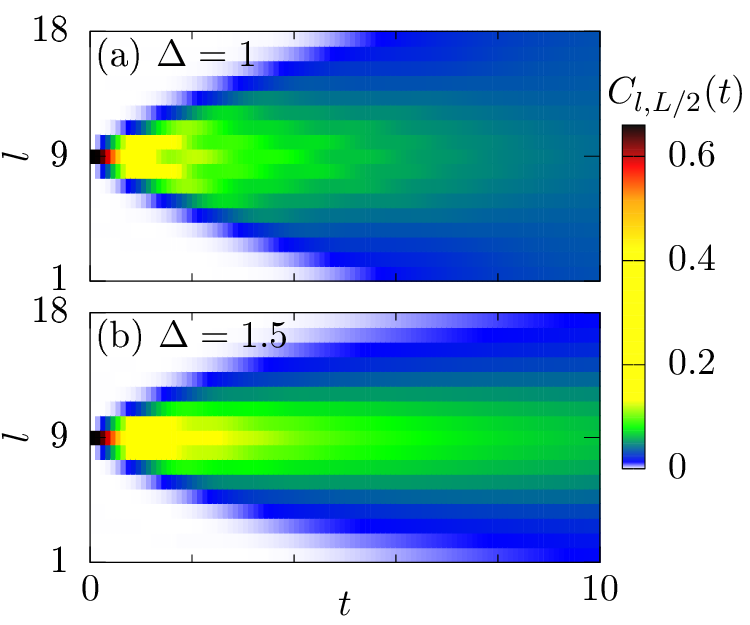}
  \caption{(Color online) Broadening of density profiles $C_{l,L/2}(t)$ at 
  infinite temperature $\beta = 0$ for (a) $\Delta = 1$ and 
  (b) $\Delta = 1.5$. We have $L = 18$ in both cases.}
  \label{Fig3}
\end{figure}

From QMC simulations, the real-space correlations $C_{l,l'}(\tau)$ are obtained 
in imaginary time $\tau$,
\begin{eqnarray}
C_{l,l'}(\tau)=\left\langle \sum_{m=0}^{M} \binom{M}{m} 
\left(\frac{\tau}{\beta}\right)^m\nonumber 
\left(1-\frac{\tau}{\beta}\right)^{M-m}\nonumber \right. \\
\left. \frac{1}{M} \sum_{p=0}^{M-1} 
S_{l}^{z}(m+p)S_{l'}^{z}(p)\right\rangle_{W} 
\, ,
\label{a2}
\end{eqnarray}
where the argument of $S_l^z(p)$ refers to discrete expansion slices of the 
SSE (for details see, e.g., \cite{GrossjohannDiss}), and 
$\langle\bullet\rangle_{W}$ denotes 
the Metropolis weight of an operator string of length $M$ generated by the SSE 
of ${\cal Z}$ \cite{Sandvik1999, Syljuasen2002}.

After a Fourier transform to momentum space, cf.\ Eq.\ \eqref{Eq::LatFou}, the 
dynamical structure factor eventually results from analytic continuation to 
real frequencies based on the inversion of
\begin{equation}
C_q(\tau) = \int_{0}^{\infty} d\omega \, C_q(\omega)K(\omega,\tau) 
\, ,
\end{equation}
with a kernel $K(\omega,\tau)=(e^{-\tau\omega}+e^{-(\beta-\tau)\omega})/\pi$.
This inversion is an ill-posed problem, for which maximum entropy methods (MEM) 
have proven to be well suited. We have applied Bryan's algorithm for our 
MEM \cite{Skilling1984, Jarrell1996}. This method minimizes the functional 
$Q=\chi^{2}/2-\alpha\sigma$, with $\chi$ being the covariance of the QMC data 
with respect to the MEM trial spectrum $C_q(\omega)$. Overfitting is prevented 
by an entropy term $\sigma=\sum_{\omega} C_q(\omega) \ln[C_q(\omega) 
/m(\omega)]$.  We have used a flat default model $m(\omega)$, which is 
iteratively adjusted to match the zeroth moment of the trial spectrum. The 
optimal spectrum follows from the weighted average of $C_q(\omega)$ with the 
probability distribution $P[\alpha| C_q(\omega)]$ \cite{Skilling1984}.


\section{Results}\label{Sec::Results}
 
We now present our numerical results. First, we study dynamics for 
infinite temperatures $\beta = 0$ and vanishing disorder $W = 
0$ in Sec.\ 
\ref{Sec::Results::CH}. Then, we also consider finite temperatures $\beta > 0$ 
in Sec.\ \ref{Sec::Results::CL}, before discussing 
the effect of disorder $W > 0$ in Sec.\ \ref{Sec::Results::DL}. 
 
\subsection{Clean model at high temperatures}\label{Sec::Results::CH}
 
\subsubsection{Current dynamics}

Let us start with the discussion of current dynamics for the isotropic model 
with $\Delta = 1$. In Fig.\ \ref{Fig1}~(a), the current autocorrelation 
function $\langle j(t)j \rangle/L$ is shown for different system sizes $L = 12, 
14, 16, 18$ in a semilogarithmic plot. First of all, for the small system 
with $L = 12$, we find that the data obtained by DQT reproduce ED 
results very accurately \cite{Steinigeweg2010}. As explained in Sec.\ 
\ref{Sec::DQT}, this accuracy is expected to improve even further if $L$ is 
increased, such that the pure-state approach can be regarded as practically 
exact for all $L \geq 12$. Moreover, while the curves are converged in system 
size at short times, finite-size effects become apparent for $t \gtrsim 10$. 
For such times, one finds that $\langle j(t)j\rangle/L$ decays to smaller and 
smaller values for increasing $L$ (although it is difficult to estimate the $L 
\to \infty$ value based on the system sizes numerically available).  
\begin{figure}[tb]
  \centering 
  \includegraphics[width=0.9\columnwidth]{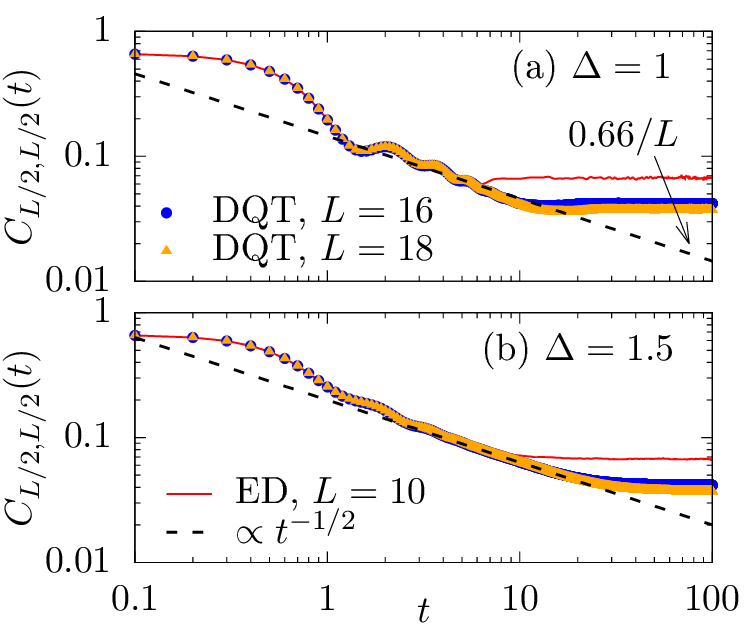}
  \caption{(Color online) (a) Equal-site correlation $C_{L/2,L/2}(t)$ for 
  $\Delta = 1$ and system sizes $L = 10$ (ED), $L = 16, 18$ (DQT), in 
a logarithmic plot. The dashed line indicates power-law decay $\propto 
1/\sqrt{t}$. The constant long-time value scales as $\tfrac{2}{3}/L$ 
\cite{Wurtz2018, Richter2018_2}. (b) Same data as in (a), but now for the 
larger anisotropy $\Delta = 1.5$. We have $\beta = 0$ in both cases.}
  \label{Fig4}
\end{figure}

Next, Fig.\ \ref{Fig1}~(b) shows the corresponding diffusion coefficient $D(t)$, 
i.e., essentially the integral over the curves shown in panel~(a). Even for the 
largest system size $L = 18$, we observe that $D(t)$ still exhibits a finite 
slope $\partial_t D(t) > 0$ and does not saturate to a constant plateau on the 
time scales shown here. It is instructive to compare these results to a 
calculation restricted to the symmetry subspace with $S^z = 0$. For this 
choice, one finds that $\langle j(t) j \rangle$ decays significantly faster, 
cf.\ Fig.\ \ref{Fig1}~(a), and correspondingly, the diffusion coefficient 
$D(t)$ is approximately constant for times $t \gtrsim 10$, cf.\ Fig.\ 
\ref{Fig1}~(b). Importantly however, the $S^z = 0$ data in Figs.\ 
\ref{Fig1}~(a), (b) are converged in system size only up to short 
times $t \lesssim 5$. Moreover, the convergence towards $L \to \infty$ is 
generally slower than a calculation in the full Hilbert space (cf.\ Ref.\ 
\cite{steinigeweg2014_2}), which can also 
be seen by comparing to results obtained by the time-dependent density matrix 
renormalization group (tDMRG) \cite{DeNardis2019, NotetDMRG}.
\begin{figure}[tb]
\centering 
\includegraphics[width=0.9\columnwidth]{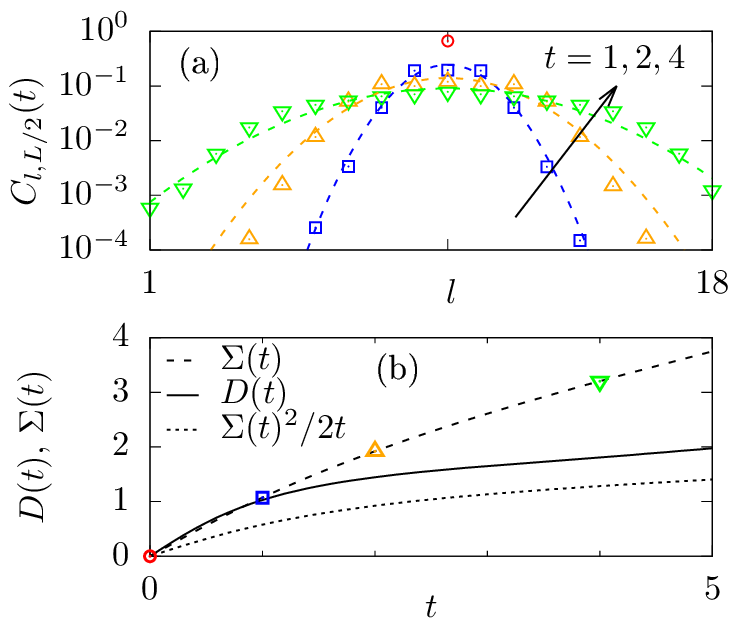}
\caption{(Color online) (a) Density profile $C_{l,L/2}(t)$ at 
fixed times $t = 0,1,2,4$. Dashed curves are Gaussian fits to the data. (b) The 
width $\Sigma(t)$, as obtained from these density profiles [symbols, Eq.\ 
\eqref{Eq::SigDense}] is compared to the width $\Sigma(t)$, as obtained from 
current autocorrelations [curve, Eq.\ \eqref{Eq::Sigt}]. The derivative $D(t)$ 
in Eq.\ \eqref{Eq::Sigt} as well as the quantity $\Sigma^2/(2t)$ are shown 
as well. The other parameters are $\Delta = 1$, $L = 18$, and $\beta = 0$.}
\label{Fig5}
\end{figure}

Eventually, Fig.\ \ref{Fig1}~(c) shows the frequency-dependent conductivity 
$\sigma(\omega)$ for the largest system size $L = 18$. We depict data for 
different frequency resolutions ${\delta \omega = \pi/10}$ and ${\delta \omega 
= \pi/20}$, i.e., two rather short cutoff times $t_\text{max}$ for the Fourier 
transform \eqref{Eq::SigOm}. However, even for the short $t_\text{max}$ chosen, 
we observe that $\sigma_\text{dc}$ strongly depends on $\delta \omega$ and 
diverges if $t_\text{max}$ is increased. Again, let us compare these results 
to a calculation in the $S^z = 0$ subsector only. In this case, the maximum of 
$\sigma(\omega)$ is shifted to a finite frequency $\omega_\text{max} \approx 
0.1$ and $\sigma(\omega)$ develops a local minimum at $\omega = 0$. Moreover, 
our DQT results are in good agreement with earlier data obtained by the 
microcanonical Lanczos method (MCLM) for $S^z = 0$ \cite{Karadamoglou2004}. 

Let us now consider a larger anisotropy $\Delta = 1.5$. Analogous to Fig.\ 
\ref{Fig1}, we present a finite-size scaling of $\langle j(t) j \rangle$ and 
$D(t)$ in Figs.\ \ref{Fig2}~(a), (b). Compared to the isotropic point, we 
find that $\langle j(t) j \rangle$ now decays to significantly smaller values. 
Moreover, as can be seen in Fig.\ \ref{Fig2}~(b), the diffusion coefficient 
$D(t)$ converges to an approximately constant and $L$-independent plateau for 
times $2 \lesssim t \lesssim 5$. In particular, this plateau persists for longer 
and longer times if $L$ is increased \cite{steinigeweg2014_2}. 

For system size $L = 16$ and $L = 18$, we 
again depict in Fig.\ \ref{Fig2}~(c) the corresponding conductivity 
$\sigma(\omega)$ for two different frequency resolutions $\delta \omega = 
\pi/30$ and $\delta \omega = \pi/50$. In contrast to $\Delta = 1$, cf.\ Fig.\ 
\ref{Fig1}~(c), we now find a well-defined dc conductivity $\sigma_\text{dc}$, 
which is practically independent of the specific $L$ and $\delta \omega$ 
chosen.  

Comparing the results presented in Figs.\ \ref{Fig1} and \ref{Fig2}, the 
dynamics of the spin current apparently exhibits qualitative differences 
between $\Delta = 1$ and $\Delta = 1.5$. On the one hand, for the isotropic 
model, there is a large discrepancy between calculations in the canonical and 
grand-canonical ensemble. On the other hand, for $\Delta = 1.5$, the 
finite-size scaling of $D(t)$ clearly suggests diffusive spin transport for 
this value of anisotropy. This is a first central result of the present paper.  

\subsubsection{Density dynamics} 
\begin{figure}[tb]
  \centering 
  \includegraphics[width=0.9\columnwidth]{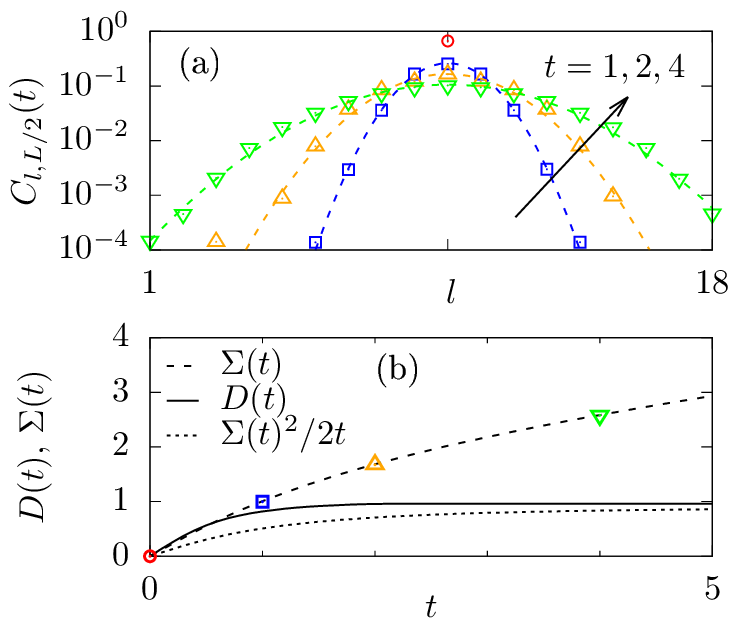}
  \caption{(Color online) Analogous data as in Fig.\ \ref{Fig5}, but now for 
  the larger anisotropy $\Delta = 1.5$. The density profiles in (a) are well 
described by Gaussians over three orders of magnitude.}
  \label{Fig6}
 \end{figure}

We now come to the discussion of the spatio-temporal correlation functions 
$C_{l,l'}(t)$. In particular, we here prepare the typical pure state 
$\ket{\tilde{\psi}(t)}$ by applying the operator $(S_{L/2}^z + 1)^{1/2}$, such 
that the expectation value of $S_l^z$ 
yields the correlation $C_{l,L/2}(t) = \langle S_l^z(t) S_{L/2}^z \rangle$, 
cf.\ Eqs.\ \eqref{Eq::Typ2} and \eqref{Eq::Psi}.
(Note that the specific value $l' = L/2$ is arbitrary due to 
periodic boundary conditions.) 

In Fig.~\ref{Fig3}, $C_{l,L/2}(t)$ is shown for $\Delta = 1, 1.5$ 
and $L = 18$ sites. As discussed in the context of Eq.~\eqref{Eq::InitProfile}, 
$C_{l,L/2}(t)$ exhibits an initial $\delta$-peak profile at $l = L/2$ which 
becomes broader for times $t > 0$. Moreover, comparing $\Delta = 1$ and $\Delta 
= 1.5$, this broadening turns out to be slower in the case of the larger 
anisotropy. 

\begin{figure}[tb]
 \centering
 \includegraphics[width = 0.9\columnwidth]{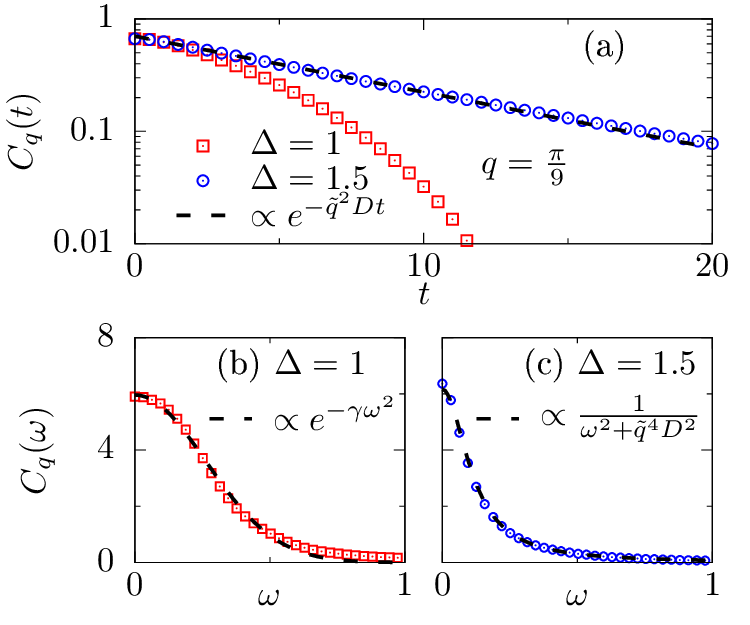}
 \caption{(Color online) Structure factors for the smallest nonzero momentum 
  $q = \pi/9$ in a chain with $L = 18$. (a) $C_q(t)$ for $\Delta = 1, 1.5$ in a 
  semilogarithmic plot. (b) $C_q(\omega)$ for $\Delta = 1$. (c) $C_q(\omega)$ 
  for $\Delta = 1.5$. The dashed lines indicate an exponential decay, as well 
  as Gaussian or Lorentzian line shapes, respectively. We have $\beta = 0$ in 
  all cases, and $\delta \omega = \pi/50$ in (b) and (c).}
 \label{Fig7}
\end{figure}
Next, let us study the decay of the central peak at $l = L/2$, i.e., the 
dynamics of the equal-site correlation function $C_{L/2,L/2}(t)$. In 
Fig.~\ref{Fig4}, $C_{L/2,L/2}(t)$ is shown for $\Delta = 1, 1.5$ 
in a logarithmic plot for different system sizes $L = 10$ (ED) and $L = 16,18$ 
(DQT). In all cases, we find that $C_{L/2,L/2}(t)$ exhibits a fast decay for 
short times $t \lesssim 1$, followed by a slower decay for $t \lesssim 
10$. In particular, for this intermediate regime, $C_{L/2,L/2}(t)$ is 
convincingly described 
by a power-law $\propto t^{-1/2}$, consistent with diffusion phenomenology, 
cf.~Eq.~\eqref{Eq::PL}. However, comparing 
$\Delta = 1$ and $\Delta = 1.5$, it appears that this power-law decay is 
cleaner for larger $\Delta$. (See also \cite{Capponi2019} for further data  
at $\Delta = 1$ and temperatures $\beta \geq 0$.) 
Eventually, for even longer times $t \gtrsim 10$, $C_{L/2,L/2}(t)$ saturates at 
a constant plateau which is related to the conservation of total magnetization 
(the plateau scales as $\propto 1/L$), cf.\ Refs.\ 
\cite{Wurtz2018, Richter2018_2}.

In order to analyze the difference between $\Delta = 1$ and $\Delta = 1.5$ in 
more detail, Figs.\ \ref{Fig5}~(a) and \ref{Fig6}~(a) show cuts of 
$C_{l,L/2}(t)$ at fixed times $t = 0,1,2,4$ for both values of $\Delta$. In the 
case of $\Delta = 1$ [Fig.~\ref{Fig5}~(a)], we find that the density profiles 
exhibit a flat region in the center of the chain which cannot be captured by 
Gaussian fits. Moreover, as shown in Fig.\ \ref{Fig5}~(b), while the widths 
$\Sigma(t)$ of these profiles necessarily agree with a calculation via current 
autocorrelations, cf.~Eq.~\eqref{Eq::Sigt}, the nonconstant $D(t)$ is 
inconsistent with diffusion. (As a consequence, $\Sigma(t) \propto 
t^{\alpha}$ with $\alpha > 1/2$ and $\Sigma(t)^2/(2t)$ is 
nonconstant.) In contrast, for $\Delta = 1.5$ 
[Fig.~\ref{Fig6}~(a)], we find that 
$C_{l,L/2}(t)$ is well described by Gaussians over roughly three orders of 
magnitude for all times shown here. These Gaussian profiles, in combination 
with the constant plateau of $D(t)$, the corresponding square-root 
growth of $\Sigma(t)$, and the saturation of  
$\Sigma(t)^2/(2t)$ in Fig.\ \ref{Fig6}~(b), are clear signatures of 
diffusion for this anisotropy. This is another important result of the present 
paper. Note that a very similar behavior, both for $\Delta = 1$ and $\Delta = 
1.5$, has been found also for $S = 1/2$ \cite{Steinigeweg2017, 
Richter2018_1}.  
 
Next, we consider correlations in momentum space. In Fig.\ \ref{Fig7}~(a), the 
intermediate structure factor $C_q(t)$ is shown in a semilogarithmic plot for a 
single system size $L = 18$ and the smallest nonzero momentum $q = \pi/9$ 
available. On the one hand, for $\Delta = 1.5$, we find that $C_q(t)$ exhibits 
a clean exponential decay with the decay rate $-\tilde{q}^2 D$, cf.\ 
Eq.~\eqref{Eq::Expo}. In particular, let us stress that the dashed line in 
Fig.\ \ref{Fig7}~(a) is no fit, but takes into account the actual value of $q$ 
and the diffusion coefficient $D \approx 0.95$ (cf.\ Refs.\ 
\cite{Steinigeweg2010, Huber2012}), as extracted from the constant 
plateau in Fig.\ \ref{Fig2}~(b). On the other hand, for $\Delta = 1$, $C_q(t)$ 
decays rather quickly and we are unable to detect an exponentially decaying 
mode for the $q$ values available. This difference between the two anisotropies 
also carries over to the frequency 
domain. In Figs.\ \ref{Fig7}~(b), (c) the dynamical structure 
factor $C_{q=\pi/9}(\omega)$ is shown for $\Delta = 1$ and $\Delta = 
1.5$, respectively. While for $\Delta = 1$, $C_q(\omega)$ is very similar to a 
Gaussian, we observe a pronounced Lorentzian line shape in the case of $\Delta 
= 1.5$, as expected for a diffusive process [cf.\ Eq.\ \eqref{Eq::Lorentz}].
\begin{figure}[tb]
  \centering
  \includegraphics[width = 0.95\columnwidth]{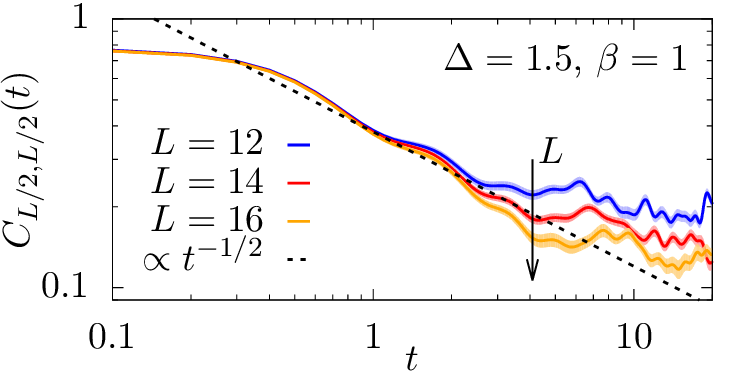}
  \caption{(Color online) Equal-site correlation $C_{L/2,L/2}(t)$ for $\Delta = 
  1.5$ and $L = 12,14,16$ at the finite temperature $\beta = 1$. Data are 
  averaged over $N_S = 50$ random initial states and the shaded area indicates 
  the standard deviation of the mean. The dashed line shows the prediction 
  $\propto 1/\sqrt{t}$ from diffusion phenomenology.}
  \label{Fig8}
\end{figure}

\subsubsection{Intermediate summary}

Based on the numerical evidence presented in Figs.\ \ref{Fig1} - \ref{Fig7}, 
high-temperature spin dynamics in the $S = 1$ XXZ chain appears to be strongly 
dependent on the value of anisotropy. On the one hand, for $\Delta = 1.5$ 
numerous signatures of genuine spin diffusion can be observed. On the other 
hand, for $\Delta = 1$, these signatures are either less pronounced or entirely 
absent. While our numerical results cannot rule out that diffusion will 
eventually emerge also for $\Delta = 1$ asymptotically at long times and 
larger $L$, they might suggest that high-temperature spin transport in the 
isotropic $S = 1$ Heisenberg chain is superdiffusive, analogous to the case of 
$S = 1/2$ \cite{Ljubotina2017, Gopalakrishnan2018, Richter2019}, and consistent 
with recent results in Ref.\ \cite{DeNardis2019}.  
 
\subsection{Clean model at lower temperatures}\label{Sec::Results::CL}
\begin{figure}[tb]
  \centering
  \includegraphics[width = 1\columnwidth]{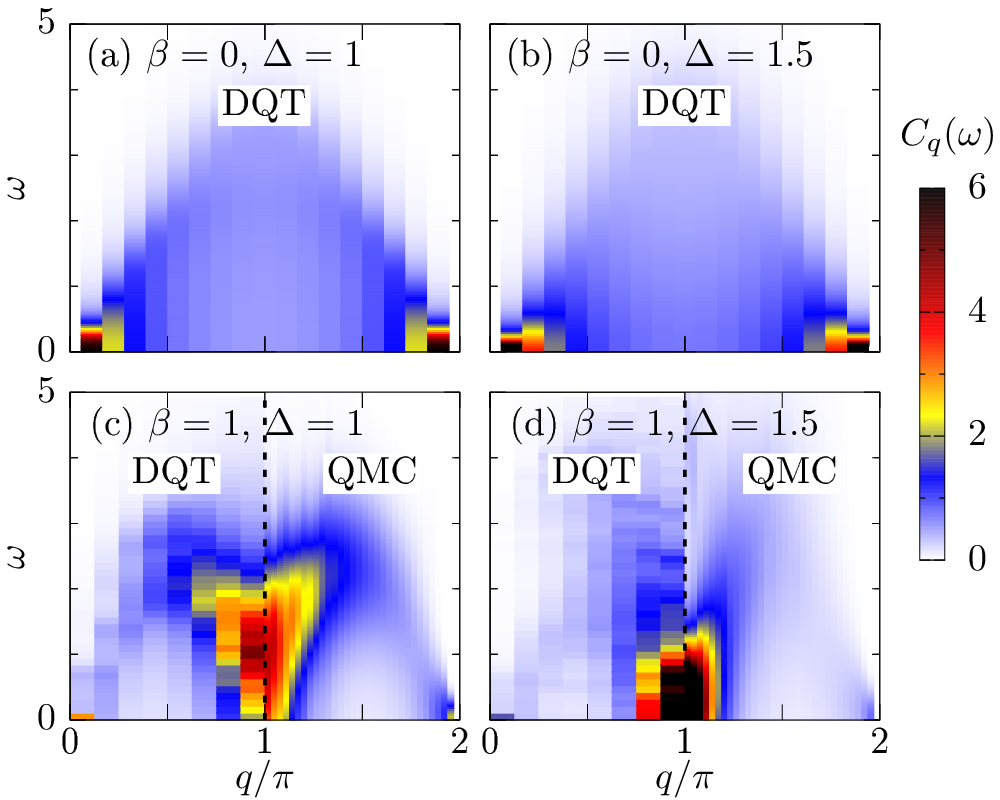}
  \caption{(Color online) (a), (b) Dynamical structure factor $C_q(\omega)$ 
at infinite temperature $\beta = 0$ for $\Delta = 1$ and $\Delta = 1.5$, 
obtained by DQT for system size $L = 18$. (c), (d) 
$C_q(\omega)$ at $\beta = 1$. Data obtained by DQT ($N_S = 50$) for $L = 16$ 
and $q < \pi$ are compared to QMC simulations for $L = 64$ at $q > \pi$.}
  \label{Fig9}
\end{figure}

In Sec.\ \ref{Sec::Results::CH}, we have unveiled clear signatures of 
high-temperature spin diffusion in the easy-axis regime ${\Delta = 1.5}$. 
Focusing on density dynamics, let us study if such signatures can be found for 
finite temperatures $\beta > 0$ as well. (See also Refs.\ \cite{Sachdev1997, 
Fujimoto1999, Konik2003, Pires2009} for transport studies low $T$.)
 
Figure \ref{Fig8} shows the equal-site correlation 
$C_{L/2,L/2}(t)$ for anisotropic chains with $\Delta = 1.5$ and $L = 12,14,16$ 
at the moderate temperature $\beta = 1$. The data are averaged over $N_S = 50$ 
random initial states in order to account for the larger statistical error of 
the typicality approximation at $\beta > 0$, cf.\ Sec.\ \ref{Sec::Av}, and the 
shaded area indicates the standard deviation of the mean. Remarkably, we are 
able to detect an intermediate time window $1 \lesssim t \lesssim 5$, where the 
decay of $C_{L/2,L/2}(t)$ is approximately described by $\propto t^{-1/2}$. 
Even though this scaling is certainly less convincing compared to the 
infinite-temperature case shown in Fig.\ \ref{Fig4}~(b), it suggests that 
diffusion might occur also at finite temperatures $T \sim J$.
\begin{figure}[tb]
 \centering
 \includegraphics[width = 1\columnwidth]{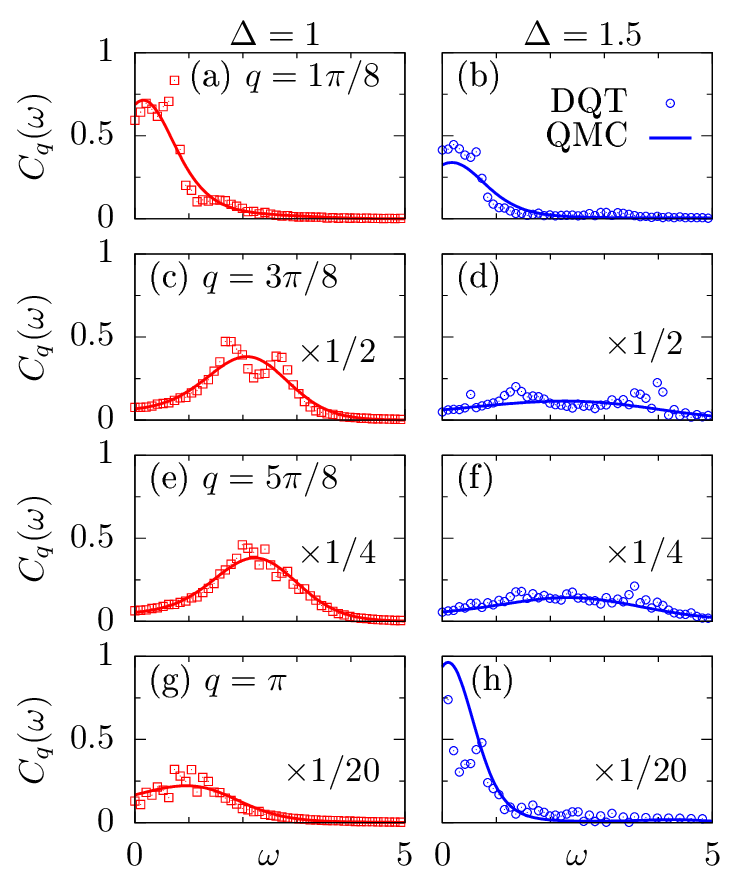}
 \caption{(Color online) Dynamical structure factor $C_q(\omega)$ at 
finite temperature $\beta = 1$ and various momenta $q$, both for $\Delta = 1$ 
(left) and $\Delta = 1.5$ (right). The data are obtained by DQT (symbols) and 
QMC (curves) for chains with $L = 16$. Note that the data in 
panels (c) - (h) have been multiplied by a factor for better visibility.}
 \label{Fig10}
\end{figure}

Next, Fig.\ \ref{Fig9} shows a contour plot of the dynamical structure factor 
$C_q(\omega)$ for all four possible combinations of $\beta = 0,1$ and 
$\Delta = 1,1.5$. On the one hand, for $\beta = 0$ [Figs.\ \ref{Fig9}~(a), 
(b)], the data are obtained by means of DQT for chains with $L = 18$. We find 
that $C_q(\omega)$ exhibits a broad excitation continuum in the center of the 
Brillouin zone extending up to $\omega \lesssim 5$, as well as distinct 
(diffusion) poles for small wave numbers $q \to 0$ [which have been 
discussed in detail in the context of Figs.\ \ref{Fig7}~(b), (c)].
On the other hand, for $\beta = 1$ [Figs.\ \ref{Fig9}~(c), (d)], we compare 
data obtained by DQT for $L = 16$ to QMC simulations for significantly larger 
systems with $L = 64$ sites. One clearly observes that the lowering of the
temperature leads to a redistribution of spectral weight. Specifically, we 
find increased intensity around $q = \pi$, which is most pronounced for $\Delta 
= 1.5$. Correspondingly, the spectral weight of the original poles for momenta 
$q \to 0$ is reduced at $\beta = 1$. Moreover, considering the big difference 
in system size, the agreement between DQT and QMC is quite convincing. For 
a thorough discussion of $C_q(\omega)$ at $\Delta = 1$ and lower temperatures 
$T \ll J$ see, e.g., Refs.\ \cite{GrossjohannDiss, Becker2017}.

For a more detailed comparison between DQT and QMC as well as between $\Delta = 
1$ and $\Delta = 1.5$, we depict cuts of $C_q(\omega)$ at $\beta = 1$ for 
various momenta $q$ in Figs.\ \ref{Fig10}~(a) - (h). In particular, DQT and 
QMC data are compared for the same chain length $L =16$. For all combinations 
of $q$ and $\Delta$ shown here, we find that DQT and QMC agree very well. While 
the DQT data are somewhat noisy due to the finite chain length, the QMC curves 
are naturally very smooth. Moreover, due to difficulties within the analytic 
continuation, it is hard to resolve certain fine structure of $C_q(\omega)$ in 
QMC simulations such as, e.g., the double peak in Fig.\ \ref{Fig10}~(c) (see 
also the discussion in \cite{Becker2017}). 
\begin{figure}[tb]
  \centering 
  \includegraphics[width=0.9\columnwidth]{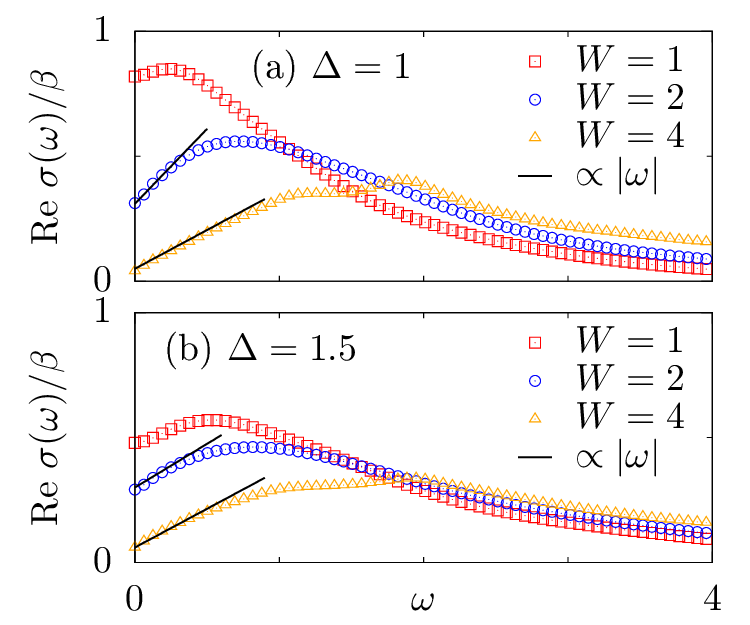}
  \caption{(Color online) Conductivity $\sigma(\omega)$ for disorder 
  $W = 1,2,4$ and anisotropies (a) $\Delta = 1$, (b) $\Delta = 1.5$. 
In the low-$\omega$ regime, the conductivity is well described 
by power laws, $\text{Re}\ \sigma(\omega) \approx \sigma_\text{dc} + 
a|\omega|^\alpha$, with $\alpha = 1$, cf.\ Refs.\ \cite{Steinigeweg2016, 
Prelovsek2017}. We have $\beta = 0$, $L = 16$, and $N = 100$ in all cases.}
  \label{Fig11}
\end{figure}

For the smallest momentum $q = \pi/8$ available [Figs.\ \ref{Fig10}~(a), (b)], 
we find that $C_q(\omega)$ behaves qualitatively similar for $\Delta = 1$ and 
$\Delta = 1.5$. Specifically, in both cases $C_q(\omega)$ has a pole at 
$\omega \approx 0$, reminiscent of the $\beta = 0$ results discussed in Figs.\ 
\ref{Fig7}~(b), (c). Moreover, the maximum of $C_q(\omega)$ seems to be 
slightly shifted to finite frequencies $\omega > 0$, although this can be a 
finite-size effect. Next, for momenta $q \approx \pi/2$ [Figs.\ 
\ref{Fig10}~(c)-(f)], we find that $C_q(\omega)$ exhibits a distinct 
excitation mode in the isotropic case, whereas the spectrum for $\Delta = 1.5$ 
is rather flat. Furthermore, as shown in Figs.\ \ref{Fig10}~(g), (h), there 
is high spectral weight at $q = \pi$, and $C_q(\omega)$ has a pronounced peak 
at $\omega \approx 0$ for $\Delta = 1.5$ (consistent with a N\'{e}el phase for 
this $\Delta$ at low $T$ \cite{Chen2003}). 

\subsection{Disordered model at high temperatures}\label{Sec::Results::DL}

Eventually, let us study the effect of disorder on the spin dynamics, focusing 
on high temperatures $\beta  = 0$. Due to the additional numerical costs caused 
by the necessity to average over different disorder realizations, we here 
restrict ourselves to a maximum system size of $L = 16$. 

Analogous to Sec.\ \ref{Sec::Results::CH}, we start our discussion with current 
dynamics. In Fig.\ \ref{Fig11}, the conductivity $\sigma(\omega)$ is shown for 
$\Delta = 1, 1.5$ and various values of disorder $W = 1, 2, 4$. 
Overall, we find a very similar behavior for both values of the exchange 
anisotropy. Specifically, for all cases shown here, one observes that 
$\sigma(\omega)$ has a well-defined dc conductivity $\sigma_\text{dc}$, which 
decreases for larger $W$. Moreover, one finds that the maximum of 
the conductivity $\sigma_\text{max} > \sigma_\text{dc}$ is shifted to larger 
and larger $\omega$ if disorder is increased. Furthermore, for 
$W = 2,4$ and low frequencies $\omega \lesssim 1$, the conductivity is well 
described by a power law, $\text{Re}\ \sigma(\omega) \approx \sigma_\text{dc} + 
a|\omega|^\alpha$, with $\alpha = 1$ \cite{Karahalios2009, Gopalakrishnan2015}. 
Note that qualitatively similar 
results for $S=1/2$ can be found in Refs.\ \cite{Steinigeweg2016, 
Prelovsek2017}.
\begin{figure}[tb]
  \centering 
  \includegraphics[width=0.9\columnwidth]{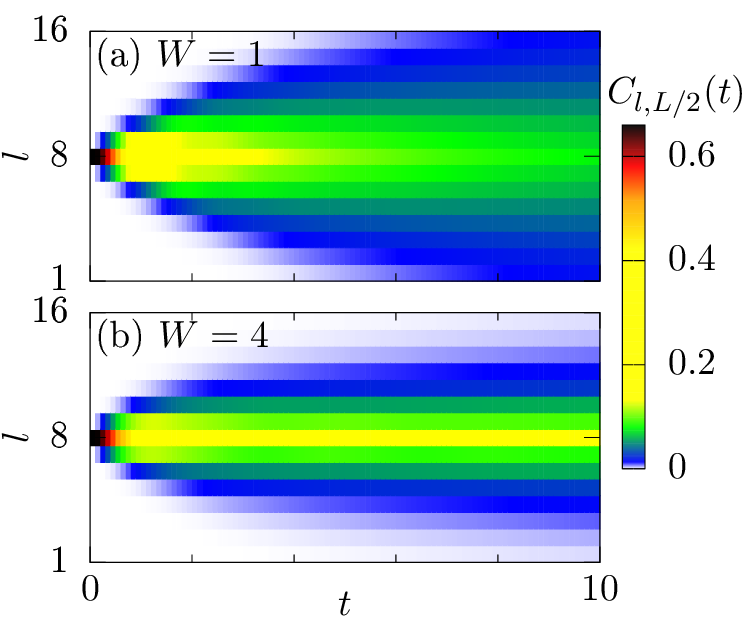}
  \caption{(Color online) Broadening of density profiles $C_{l,L/2}(t)$ for 
  different values of disorder (a) $W = 1$, (b) $W = 4$. We have $\Delta = 
  1.5$, $\beta = 0$, $L = 16$, and $N = 100$ in both cases.}
  \label{Fig12}
\end{figure}

Next, let us also discuss the dynamics of spatio-temporal correlations 
$C_{l,L/2}(t)$ in the presence of disorder. We here particularly focus on the 
easy-axis regime $\Delta = 1.5$. For this value of $\Delta$, we have unveiled 
various signatures of diffusion in the disorder-free case $W = 0$, cf.\ Figs.\ 
\ref{Fig3} - \ref{Fig7}. These data now serve as a benchmark for the study of 
$W > 0$. In Figs.\ \ref{Fig12}~(a) and (b), contour plots of $C_{l,L/2}(t)$ 
are shown for $W = 1$ and $W = 4$, respectively. Analogous to Fig.\ \ref{Fig3}, 
$C_{l,L/2}(t)$ initially exhibits a $\delta$ peak at $t = 0$, which 
broadens for times $t > 0$. However, this broadening becomes slower if $W$ is 
increased, with more weight remaining close to the center of the chain.  

\begin{figure}[tb]
  \centering 
  \includegraphics[width=0.95\columnwidth]{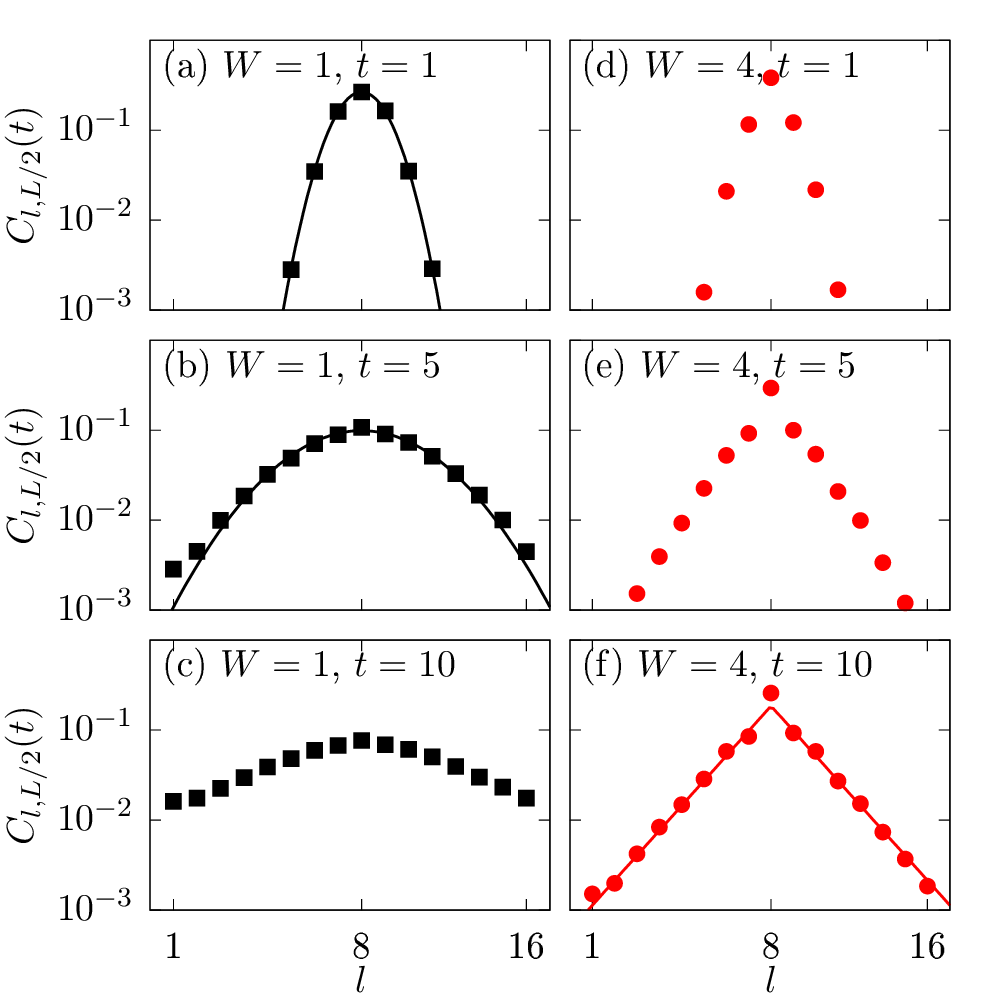}
  \caption{(Color online) $C_{l,L/2}(t)$ at fixed times $t = 1, 5, 10$ for 
  $W = 1$ [(a) - (c)] and $W = 4$ [(d) - (f)] in a semilogarithmic plot. Curves
  indicate Gaussian or exponential fits. We have $\Delta = 1.5$, $\beta = 0$, 
  $L = 16$, and $N = 100$ in all cases.}
  \label{Fig13}
\end{figure}

For a more detailed analysis, Fig.\ \ref{Fig13} shows cuts of $C_{l,L/2}(t)$ at 
fixed times $t = 1, 5, 10$ for both, weak disorder $W = 1$ [(a) - 
(c)] and 
stronger disorder $W = 4$ [(d) - (f)]. On the one hand, for $W = 1$, we find 
that $C_{l,L/2}(t)$ is again in good agreement with a Gaussian profile
at short times $t = 1$. In comparison with the $W = 0$ data shown 
in Fig.\ \ref{Fig6}~(a), this agreement becomes slightly less convincing for 
later times $t = 5$ [Fig.\ \ref{Fig13}~(b)]. On the other hand, for $W = 4$, 
$C_{l,L/2}(t)$ exhibits a completely different behavior. For times $t = 5$ and 
$10$, 
cf.\ Figs.\ \ref{Fig13}~(e), (f), the profiles are not described by Gaussians 
anymore, but are rather of triangular shape in the semilogarithmic plot used. 
Such exponentially decaying tails are clearly inconsistent with diffusion and 
might suggest the presence of a nondiffusive regime 
\cite{Richter2018_2, Bera2017, Weiner2019}.

\begin{figure}[b]
  \centering 
  \includegraphics[width=0.9\columnwidth]{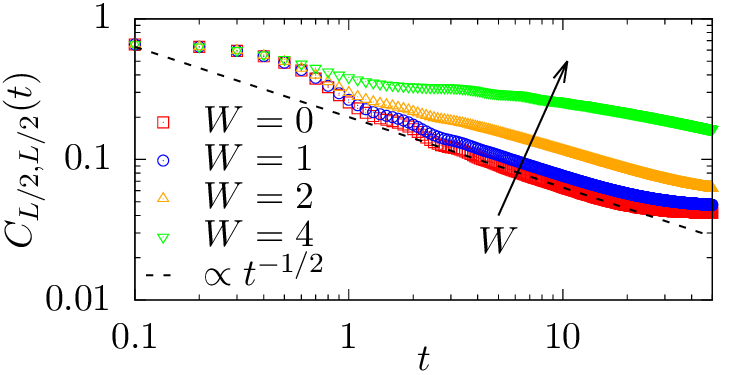}
  \caption{(Color online) $C_{L/2,L/2}(t)$ for disorder $W = 0, 1, 2, 4$ 
  in a logarithmic plot. We have $\Delta = 1.5$, $\beta = 0$, $L = 16$, and 
$N = 100$ in all cases.}
  \label{Fig14}
\end{figure}

\begin{figure}[tb]
  \centering 
  \includegraphics[width=0.95\columnwidth]{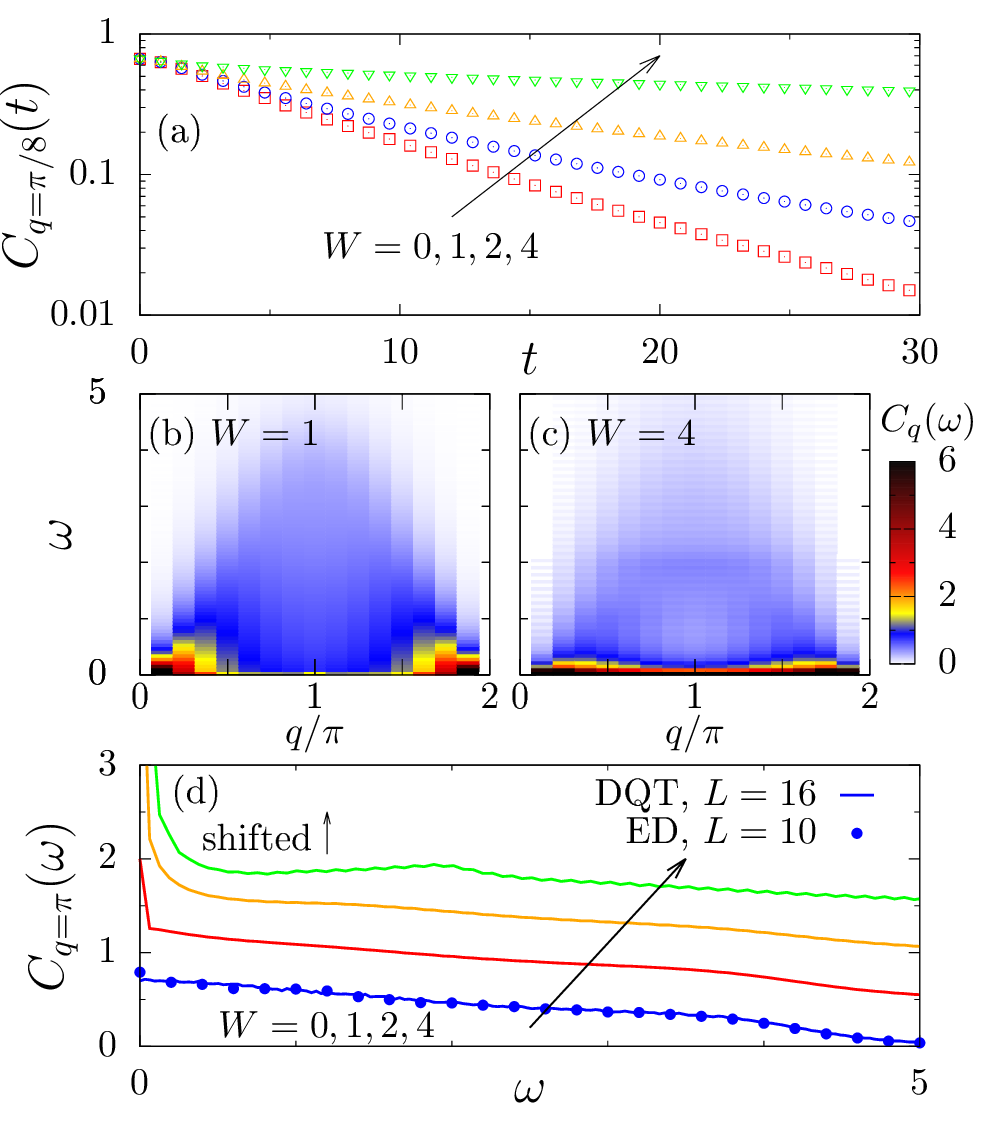}
  \caption{(Color online) (a) $C_q(t)$ at momentum $q = \pi/8$ for disorder $W 
= 0, 1, 2, 4$ in a semilogarithmic plot. (b), (c) Contour plots of 
$C_q(\omega)$ for disorder $W = 1$ and $W = 4$, respectively. (d) $C_q(\omega)$ 
at momentum $q = \pi$ for disorder $W = 0, 1,2,4$. For $W = 0$, we additionally 
compare to ED ($L = 10$). The curves for $W > 0$ are shifted by constant 
offsets in order to improve the visibility. We have $\Delta = 1.5$, $\beta = 
0$, $L = 16$, and $N = 100$ in all cases.}
  \label{Fig15}
\end{figure}

Furthermore, the equal-site correlation function $C_{L/2,L/2}(t)$ is shown in 
Fig.\ \ref{Fig14} for $W = 0, 1, 2, 4$ in a logarithmic plot. While the curves 
for $W = 0$ and $W = 1$ are still very similar to each other, we find that 
$C_{L/2,L/2}(t)$ decays slowly for strong disorder $W = 4$ and is inconsistent 
with $\propto t^{-1/2}$. (For a study of the Fourier transform of 
$C_{L/2,L/2}(t)$ in disordered spin-$1/2$ models, see \cite{Serbyn2017})
Figure \ref{Fig15}~(a) shows the intermediate structure factor 
$C_q(t)$ for the smallest nonzero momentum $q = \pi/8$. For increasing $W$, we 
find that the slope of $C_q(t)$ becomes more and more flat, which can be 
interpreted as a shrinking of the diffusion constant. In particular, for 
strong disorder $W = 4$, $C_q(t)$ essentially does not decay at all on the 
time scales depicted. 
The nondecaying behavior of $C_q(t)$ is also reflected in its Fourier transform 
$C_q(\omega)$, which is shown in Figs.\ \ref{Fig15}~(b), (c) in terms of 
a contour plot for $W = 1$ and $W = 4$, respectively. While $C_q(\omega)$ still 
exhibits a broad excitation continuum in the center of the Brillouin zone [cf.\ 
Fig.\ \ref{Fig9}~(b) for $W= 0$], we find that $C_q(\omega)$ additionally 
develops a high contribution at $\omega = 0$ if $W$ is increased.
On the one hand, for $W = 1$, this peaked structure is pronounced for 
$q \to 0$, e.g., due to diffusion. On the other hand, for strong disorder $W = 
4$, the high contribution in $C_q(\omega)$ can be clearly identified for 
all momenta $q$ in the Brillouin zone. Note that a very 
similar behavior has also been observed in the case of spin $S = 1/2$ 
\cite{Prelovsek2017}. 
To illustrate the development of this high contribution, Fig.\ \ref{Fig15}~(d) 
shows $C_q(\omega)$ for $W = 0,1,2,4$ at the fixed momentum $q = \pi$. Note 
that the data for different values of $W$ are artificially shifted in the 
vertical direction to improve visibility. For all values of disorder shown 
here, we find that $C_{q=\pi}(\omega)$ has an almost featureless shape for 
finite $\omega$ and roughly extends up to $\omega \lesssim 5$. However, one can 
clearly observe that the high contribution at $\omega \approx 0$ becomes more 
pronounced for increasing $W$. 

The numerical data presented in Figs.\ \ref{Fig11} - \ref{Fig15} suggest that 
the spin-$1$ XXZ chain undergoes a transition between a diffusive regime and a 
nondiffusive phase for sufficiently strong disorder. 
%

 
\section{Conclusion}\label{Sec::Conclusion}

To summarize, we have studied the magnetization dynamics in the 
one-dimensional $S=1$ XXZ model for various anisotropies and temperatures, as 
well as in the presence of quenched disorder induced by a random magnetic 
field.  

As a main result, we unveiled that high-temperature spin transport is diffusive 
in the easy-axis regime for strong exchange anisotropies. This finding was 
based on the combination of numerous signatures, such as (i) Gaussian spreading 
of correlations, (ii) a time-independent diffusion coefficient, (iii) power-law 
decay of equal-site correlations, (iv) exponentially decaying long-wavelength 
modes, and (v) Lorentzian line shapes of the dynamical structure factor. 
Besides, we provided evidence that some of these signatures are not 
exclusively restricted to the infinite-temperature limit, but can persist at 
lower temperatures as well. For these lower temperatures, we particularly found 
a very good agreement between the pure-state typicality approach and additional 
quantum Monte Carlo simulations. 
 
In contrast, we demonstrated that a proper analysis of magnetization dynamics 
is 
considerably more delicate for the isotropic case $\Delta = 1$. Specifically, 
we found that even for the largest system sizes amenable to our numerical 
approach, the signatures (i) - (v) are either less pronounced or entirely 
absent. Therefore, our numerical analysis suggests that high-temperature spin
transport might be superdiffusive in the $S = 1$ Heisenberg chain despite the 
nonintegrability of the model. This finding is consistent with recent results 
in Ref.\ \cite{DeNardis2019}.

Eventually, upon introducing a random on-site magnetic field, we observed a 
breakdown of diffusion and distinctly slower dynamics. Moreover, our 
results exhibit qualitative similarities to disordered spin-$1/2$ chains 
and might be consistent with the presence of a nondiffusive 
regime.

Promising directions of research include, e.g., the application of the 
pure-state approach to spin dynamics for $S \geq 1$ at finite and infinite 
temperature. In particular, a more detailed analysis of a 
putative transition to a many-body localized phase in models with $S 
\geq 1$ is an interesting avenue of future work. 
 
\subsection*{Acknowledgements}

This work has been funded by the Deutsche 
Forschungsgemeinschaft (DFG) - Grants No. 397067869 (STE 2243/3-1), 
No. 355031190 - within the DFG Research Unit FOR 2692.
N. C. and W. B. acknowledge financial support from ``Nieders\"achsisches 
Vorab'' through ``Quantum- and Nano-Metrology (QUANOMET)'' initiative within the 
project NP-2. 

\appendix 

\section{Typicality relation}\label{Sec::Appendix::Typ}

Let us briefly derive the typicality relation given in Eq.\ 
\eqref{Eq::Typ2} of the main text. To this end, we start with a correlation 
function at formally infinite temperature,  
\begin{align}
  \frac{\text{Tr}[S_l^z(t) (S_{l'}^z + 1)]}{d} 
  &= \frac{\text{Tr}[S_l^z(t) S_{l'}^z]}{d} + \frac{\text{Tr}[S_l^z(t)]}{d} \\
 					      &= C_{l,l'}(t)\ , 
 \end{align}
where we have used that $\text{Tr}[S_l^z] = 0$. Thus, the expression 
$\text{Tr}[S_l^z(t) 
(S_{l'}^z + 1)]/d$ is equivalent to the correlation function $C_{l,l'}(t)$ from 
Eq.\ \eqref{Eq::Cll}. Exploiting this fact, we can now write
 \begin{align}
  \frac{\text{Tr}[S_l^z(t) (S_{l'}^z+1)]}{d} 
  &= \frac{\text{Tr}[\sqrt{S_{l'}^z+1} S_l^z(t) \sqrt{S_{l'}^z+1}]}{d} \\
 &\approx \frac{\bra{\varphi} \sqrt{S_{l'}^z+1} S_l^z(t) \sqrt{S_{l'}^z+1} 
\ket{\varphi}}{\braket{\varphi|\varphi}}  \nonumber\\
 &= \bra{\tilde{\psi}(t)} S_l^z \ket{\tilde{\psi}(t)}\ ,  \label{Eq::Eq2}
 \end{align}
where we have used the cyclic invariance of the trace and the definition 
of the pure state $\ket{\tilde{\psi}(t)} = 
e^{-i\mathcal{H}t} \ket{\tilde{\psi}(0)}$, cf.\ Eq.\ \eqref{Eq::Psi}. Note that 
the statistical error $\epsilon$ of the typicality approximation has been 
dropped for clarity in Eq.\ \eqref{Eq::Eq2}. 

%

\end{document}